\documentclass[%
 aip,
 amsmath,amssymb,
 reprint,%
]{revtex4-1}

\usepackage{graphicx}
\usepackage{dcolumn}
\usepackage{bm}
\usepackage{comment}
\usepackage{braket}
\usepackage[dvipsnames]{xcolor}
 
\definecolor{aqua}{rgb}{0 0.2 0.7}

\usepackage[utf8]{inputenc}
\usepackage[T1]{fontenc}
\usepackage{mathptmx}
\usepackage[backref=true]{hyperref}
\hypersetup{
    colorlinks=true,
    linkcolor=aqua,
    urlcolor=aqua,
    citecolor=aqua,
}
\newcommand\Tstrut{\rule{0pt}{2.6ex}}         %
\newcommand\Bstrut{\rule[-0.9ex]{0pt}{0pt}}   %

\begin{document}
\preprint{AIP/123-QED}
\title{Mechanical Purcell Filters for Microwave Quantum Machines}

\author{Agnetta Y. Cleland}
\email{agnetta@stanford.edu}
\author{Marek Pechal}%
\author{Pieter-Jan C. Stas}
\author{Christopher J. Sarabalis}
\author{Amir H. Safavi-Naeini}
\email{safavi@stanford.edu}
\affiliation{Department of Applied Physics and Ginzton Laboratory, Stanford University\\ 348 Via Pueblo Mall, Stanford, California 94305, USA}

\date{\today}

\begin{abstract}
In circuit quantum electrodynamics, measuring the state of a superconducting qubit introduces a loss channel which can enhance spontaneous emission through the Purcell effect, thus decreasing qubit lifetime. This decay can be mitigated by performing the measurement through a Purcell filter which forbids signal propagation at the qubit transition frequency. If the filter is also well-matched at the readout cavity frequency, it will protect the qubit from decoherence channels without sacrificing measurement speed. We propose and analyze design for a mechanical Purcell filter, which we also fabricate and characterize at room temperature. The filter is comprised of an array of nanomechanical resonators in thin-film lithium niobate, connected in a ladder topology, with series and parallel resonances arranged to produce a bandpass response. The modest footprint, steep band edges, and absence of cross-talk in these filters make them a novel and appealing alternative to analogous electromagnetic versions currently used in microwave quantum machines. %
\end{abstract}

\maketitle

Quantum information processing calls for systems which are well isolated from their environment, whose states can nonetheless be measured and manipulated with precision. \cite{divincenzo} These fundamentally contradictory requirements can be satisfied by cleverly engineering devices and interactions between them. In the circuit QED platform,\cite{devoret_schoelkopf,koch_schoelkopf,neill_martinis} nonlinear superconducting circuits called qubits are used to store and process quantum information. Their internal states need to be read out rapidly and with low rates of error. An appealing approach for this is to couple the qubit circuit to an auxiliary, linear electromagnetic resonator (often called the readout cavity). Resonators have long been used to amplify emission from atoms, for instance, via Purcell enhancement.~\cite{purcell,goy_haroche, girvin} Conversely, qubit emission into the environment can be suppressed by tuning the qubit away from the resonator's frequency by many times the linewidth and interaction energy. The qubit state is then measured “dispersively” by monitoring the resonator frequency for shifts induced by changes in the qubit state~\cite{koch_schoelkopf} (although we note that alternative measurement strategies exist). \cite{non_dispersive} 

Dispersive shifts of the cavity can be measured and amplified to demonstrate extremely efficient single-shot measurements of qubits using this scheme. \cite{quantum_jumps_siddiqi, johnson_clarke_siddiqi} Nonetheless, the conflicting requirements of efficient readout and qubit isolation persist in the desired properties of the resonator. Fast, efficient readout requires strong coupling between the resonator and environment. This in turn increases the probability of qubit relaxation through the resonator in a process called Purcell decay.\cite{purcell, houck_single_photons, gambetta_blais, houck_schoelkopf} Purcell filters, often consisting of a second stage electromagnetic resonator, have been used effectively to mitigate this process.\cite{reed_schoelkopf, jeffrey_martinis, sete_martinis,bronn_chow, walter_wallraff, heinsoo_wallraff} As qubit coherence times continue to improve, the basic limit imposed by Purcell decay will become more important. In principle, progressively higher order electromagnetic filters can be incorporated, requiring progressively larger components that take up valuable space on a chip. 

We propose an alternative solution using ultra-compact, high-order microwave acoustic filters to isolate qubits from the environment and to curtail Purcell decay, using techniques adapted from the telecommunications industry. Low cross-talk and extremely small footprints make nanomechanical structures ideal for integration with emerging superconducting quantum machines.\cite{arrangoiz_PRA, satzinger_cleland,chu_schoelkopf, pechal_safavi,arrangoiz_PRX, arrangoiz_arxiv} Moreover, recent advances in fabrication and design of thin-film, high-$Q$, and strongly coupled lithium niobate (LN) devices make them well-suited for such applications.\cite{arrangoiz_PRX, arrangoiz_arxiv, sarabalis_arxiv} In this manuscript we outline the approach and present initial experimental results. 

The cavity-qubit interaction is described by the Jaynes-Cummings Hamiltonian:
\begin{equation}\label{eq:JC}
\hat H_{\text{JC}}= \hbar \omega_\text{r}\bigg(\hat a^\dagger \hat a + \frac{1}{2}\bigg) + \frac{\hbar \omega_\text{q}}{2}\hat \sigma_z + \hbar\,g \big(\hat a^\dagger \hat \sigma_- + \hat a \hat \sigma_+\big)
\end{equation}
where $\omega_\text{r}$ is the readout resonator frequency, $g$ is the qubit-resonator coupling strength, $\hat a$ annihilates photons in the resonator, and Pauli operators $\hat \sigma$ act on the qubit. This Hamiltonian can be diagonalized into a series of $n$-excitation subspaces, each spanned by $\ket{g,n}$ and $\ket{e,n-1}$. For $|g|\gg|\Delta|$, the ``qubit-like'' polariton in the first excited subspace is: $$\ket{+,1} \approx \ket{e,0}+\frac{g}{\Delta}\ket{g,1}.$$ The small but finite occupation of the resonator represented by $\ket{g,1}$ can decay to $\ket{g,0}$ through the resonator's output channel, effectively causing the atom to relax to its ground state. This loss channel increases the decay rate of the qubit by  $\gamma_\text{q} = \kappa |\langle g,0|\,\hat{a}\,|+,1\rangle|^2\approx {g^2}\kappa/{\Delta^2}$. 
 More generally, we can use Fermi's golden rule to calculate the decay of the qubit excited state $\ket{e,n}$ through the resonator:
\begin{equation} \label{eqn:Purcell}
\gamma_\text{q} = \frac{g^2\kappa}{\Delta^2+(\kappa/2)^2} \,\underset{\Delta \gg \kappa,g}{\longrightarrow}\, \frac{g^2}{\Delta^2}\kappa.
\end{equation}
 This allows us to choose $\Delta$ such that the maximum coherence time imposed by the Purcell effect $1/\gamma_\text{q}$ exceeds the qubit's $T_1$ due to other sources. With improving qubit design, fabrication, and materials processing, larger detunings $\Delta$ will be required to avoid limitation by Purcell decay. However, increasing $\Delta$ also increases the amount of time required to make a measurement, which allows for more errors to be introduced. This inconvenience, as well as other practical concerns with operating at large detunings, has led the to design and implementation of Purcell filters. These filters, previously composed of electromagnetic resonators, protect the qubit from the Purcell decay channel while maintaining the ability to do fast measurements by operating with relatively small $\Delta$.

\begin{figure}
    \centering
    \includegraphics[width = 8.5cm]{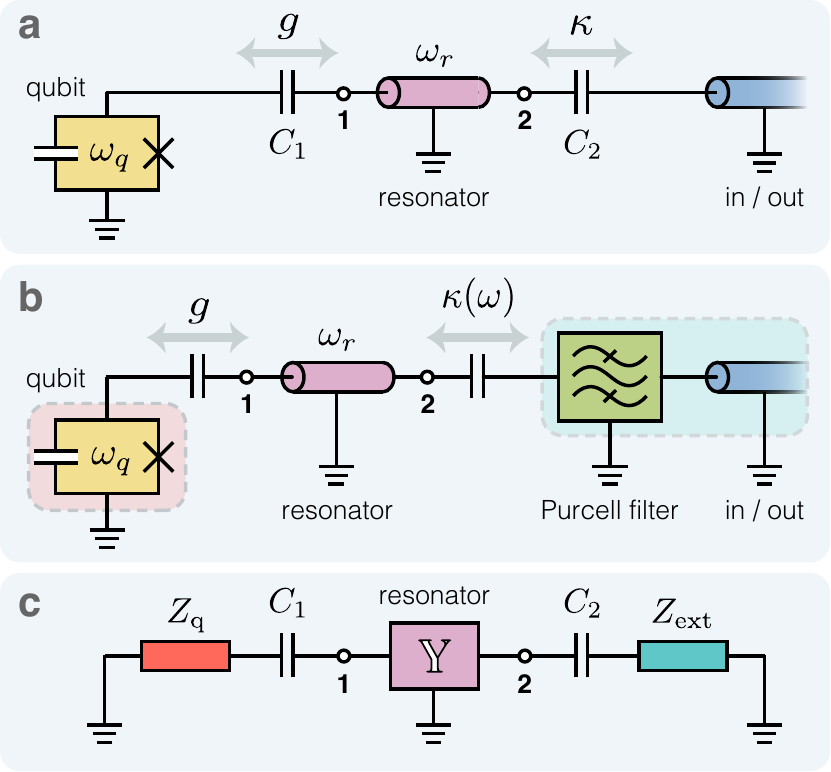}
    \caption{Dispersive qubit readout. \textbf{a.} Conventional schematic for measuring the energy state of a superconducting qubit (orange) with a transition frequency $\omega_\text{q}$. The qubit is coupled at a rate $g$ to a linear resonator (pink) with a fundamental resonance at $\omega_\text{r}$, which decays at a rate $\kappa$ to its input and output transmission line (blue). The coupling rates $g$ and $\kappa$ are controlled by the capacitances $C_1$ and $C_2$. \textbf{b.} By driving the auxiliary resonator through the passband of a Purcell filter (green), one can modify the external dissipation presented to the resonator and qubit, so that now $\kappa$ and consequently $T_1$ include a frequency dependence. \textbf{c.} To derive the filter's effect on qubit lifetime, we replace the filter and environmental dissipation with their equivalent impedance $Z_{\text{ext}}$ and the qubit with its impedance $Z_\text{q}$. The resonator response is encapsulated by its admittance matrix $\mathbb{Y}$, taken with respect to nodes 1 and 2.}
    \label{fig:readout_circ}
\end{figure}

A Purcell filter can be considered to be a bandpass filter (Fig. \ref{fig:readout_circ}), which performs an impedance transformation on the dissipative bath of the environment, through which the linear resonator can be probed. Placing the qubit frequency outside the passband where the filter presents an impedance mismatch isolates the qubit from decoherence channels of the vacuum; placing the resonator frequency within the passband allows a microwave tone to pass through unimpeded and probe the resonator frequency. The resonator can be strongly coupled to its feed line (large $\kappa$), allowing signal to pass through the cavity quickly for fast qubit state measurement, without risking a reduction in qubit lifetime. 

In design and analysis, we treat the mechanical filter as a two port microwave system connected at one end to a regular transmission line and at the other end to the readout resonator (Fig. \ref{fig:readout_circ}). The impedance $Z_\text{ext}(\omega)$ seen by the resonator captures all properties of the filter element relevant to qubit operation. In this section, we develop an understanding of how $Z_\text{ext}(\omega)$ affects qubit readout and Purcell decay.

A relevant figure of merit for a Purcell filter is the ratio of qubit lifetimes with and without the filter. In this ratio the capacitance $C_2$ is adjusted to keep the resonator linewidth fixed. We can derive this by analyzing the measurement circuitry depicted in Fig. \ref{fig:readout_circ}c, in which the capacitances $C_1$ and $C_2$ are assumed to be small. The admittance matrix for the entire system, with respect to nodes 1 and 2, is given by:
$$\mathbb{Y} + \left[\begin{array}{cc} \frac{1}{1/\text{i}\omega C_1+Z_\text{q}} & 0 \\ 0 & \frac{1}{1/\text{i}\omega C_2+Z_{\text{ext}}} \end{array}\right] \approx \mathbb{\overline Y} + \left[\begin{array}{cc} \omega^2 C_1^2 Z_\text{q} & 0 \\ 0 &  \omega^2 C_1^2 Z_{\text{ext}} \end{array}\right]$$
where we absorb the reactive part of the admittance matrix with $\mathbb Y$ into $\mathbb{\overline Y}$. We solve for the resonances by setting the determinant of the above expression to $0$. These solutions are small deviations from the uncoupled case: the resonator mode shifts to $\omega_\text{r} + \delta \omega_\text{r}$ and the qubit frequency that satisfies $Z_\text{q}(\omega_\text{q})\rightarrow\infty$ shifts to $\omega_\text{q} + \delta \omega_\text{q}$. Keeping lowest-order terms in the small capacitances we find:
\begin{align*}
  \delta\omega_{\mathrm{r}} &= + \frac{
  \omega_{\text{r}}^2 C_1^2 \mathbb{\overline{Y}}_{22}(\omega_{\mathrm{r}}) 
  }{Y'_{\mathrm{q}}(\omega_{\mathrm{q}}) \Delta\lambda} -
  \frac{\omega_{\mathrm{r}}^2 C_2^2 Z_{\mathrm{ext}}(\omega_{\mathrm{r}}) \mathbb{\overline{Y}}_{11}(\omega_{\mathrm{r}})
  }{\lambda}\\
  \delta\omega_{\mathrm{q}} &=
  - \frac{\omega_{\mathrm{q}}^2 C_1^2 \mathbb{\overline{Y}}_{22}(\omega_{\mathrm{q}})} 
  {Y'_{\mathrm{q}}(\omega_{\mathrm{q}}) \Delta\lambda} +
  \frac{\omega_{\mathrm{q}}^4 C_1^2 C_2^2 Z_{\mathrm{ext}}(\omega_{\mathrm{q}})
  \mathbb{\overline{Y}}_{22}(\omega_{\mathrm{q}}) \mathbb{\overline{Y}}_{11}(\omega_{\mathrm{q}})} 
  {Y'_{\mathrm{q}}(\omega_{\mathrm{q}}) \Delta^2\lambda^2}
\end{align*}
where we introduce $\lambda = \frac{\text{d}}{\text{d}\omega} \det \mathbb{\overline Y}(\omega)\big|_{\omega=\omega_{\text{r}}}$, $Y_\text{q}(\omega) = 1/Z_\text{q}(\omega)$, and $Y' = \text{d}Y/\text{d}\omega$. The coupling strength $g$ can be found by equating the shift of the cavity frequency to $2g^2/\Delta$ and approximating $|\Delta|\ll \omega_{\text{q,r}}$. This is given by $g^2 \approx -{\omega_{\mathrm{q}}^2 C_1^2 \mathbb{\overline{Y}}_{22}(\omega_{\mathrm{q}})}/ 
  {Y'_{\mathrm{q}}(\omega_{\mathrm{q}}) \lambda}$.
Dissipation from the external environment $Z_{\text{ext}}$ introduces a small imaginary component to the frequency shifts, from which we extract the resulting qubit and resonator linewidths $\gamma_{q},\kappa = 2\,\text{Im}\,\delta\omega_{q,r}$ to find:
\begin{equation}\label{eq:linewidths}
\gamma_\text{q} = \frac{g^2}{\Delta^2}\frac{\text{Re}\,Z_{\text{ext}}(\omega_\text{q})}{\text{Re}\,Z_{\text{ext}}(\omega_\text{r})}\,\kappa.
\end{equation}
Without a filter, when $Z_{\text{ext}}(\omega_\text{r}) =Z_{\text{ext}}(\omega_\text{q})=Z_0$, this reduces to the familiar Purcell decay rate of Eq. (\ref{eqn:Purcell}). The filter adds an extra degree of protection from spontaneous emission, a ``filter factor" which is the ratio of the resistances seen by the qubit and resonator at their respective frequencies. 

\begin{figure}
    \centering
    \includegraphics[width=8.5cm]{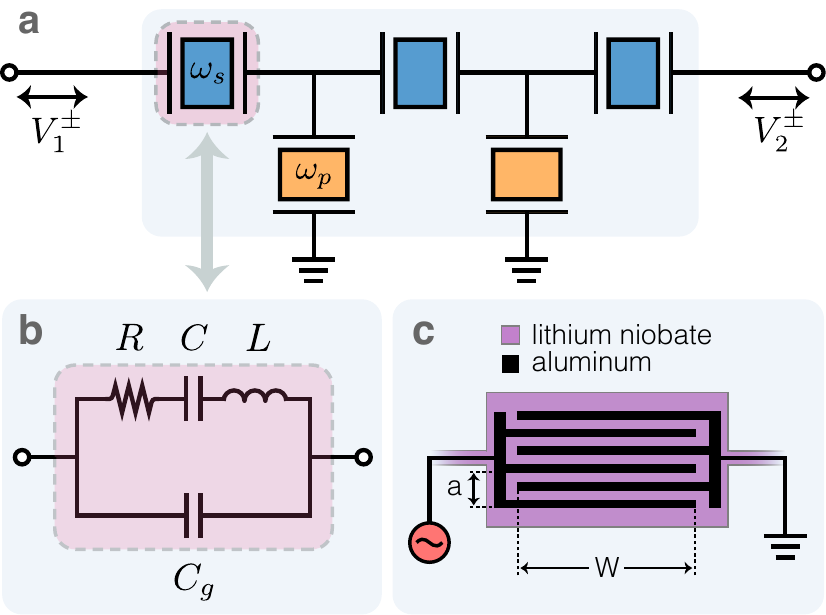}
    \caption{Nanomechanical ladder filter design. \textbf{a.} Ladder topology of order 3 in which each colored block represents an acoustic element. Three identical series resonators (blue) at $\omega_\text{s}$ are shunted by two identical parallel resonators (orange) at $\omega_\text{p}\neq \omega_s$. The highlighted element is detailed in (b) for its electrical response and (c) for its basic physical design. \textbf{b.} Butterworth-van Dyke equivalent circuit that models a single mechanical resonator. The transducer's electrostatic capacitance $C_\text{g}$ is connected in parallel with a series $RLC$ that describes acoustic activity. \textbf{c.} Diagram of a single resonator. A suspended plate of lithium niobate is patterned with aluminum interdigitated transducers of pitch $a$ and width $W$.}  
    \label{fig:MBVD}
\end{figure}

\begin{figure}
    \centering
    \includegraphics[width=8cm]{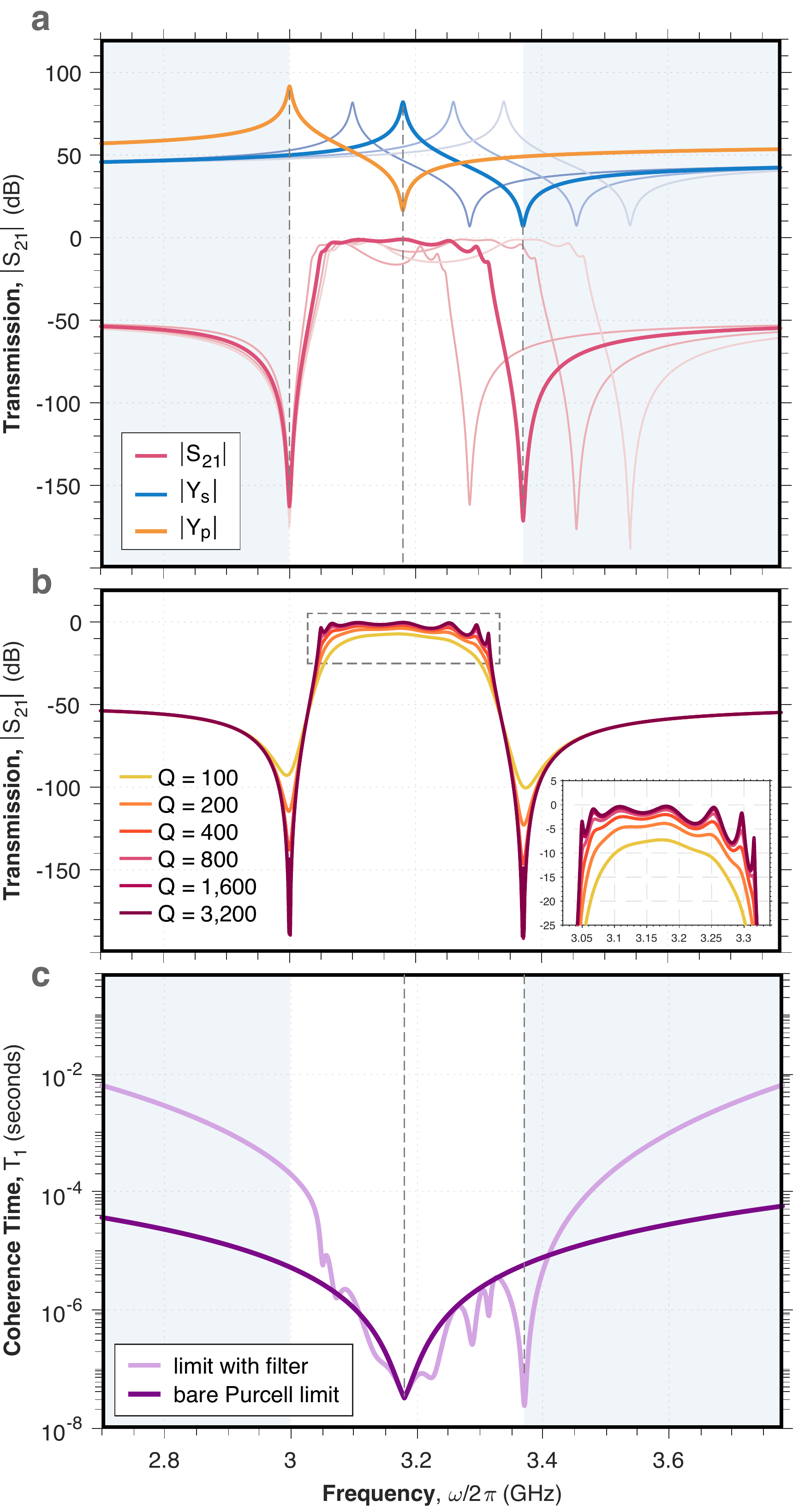}
    \caption{Circuit analysis. %
    \textbf{a.} Filter response for different detunings of the series resonance as computed by the BVD model. \textbf{b.} Filter response for $\omega_\text{p}= 2\pi \times 3.00$ GHz and $\omega_\text{s} = 2\pi \times 3.18$ GHz for different values of $Q_\text{s} = Q_\text{p}=Q$. \textbf{c.} Projected enhancement of qubit relaxation time $T_1$ with the addition of a Purcell filter, with the readout resonator frequency centered in the passband at $\omega_\text{r} = \omega_\text{s} = 2\pi \times 3.18$ GHz. This enhancement is calculated according to Eqs. \eqref{eq:linewidths} and \eqref{eq:S11_to_Zext} for the geometry corresponding to the bold traces in (a), which matches the $Q=800$ trace in (b). The sharp dip in filtered $T_1$ at 3.35 GHz corresponds to the series antiresonance of the filter.}
    \label{fig:BP_f2_Q}
\end{figure}

The mechanical Purcell filter is  based on a ladder network of  piezoelectric oscillators, inspired by methods that are ubiquitous in classical RF and telecommunications technology. \cite{morgan,hashimoto} A ladder filter electrically connects series and shunt resonators with frequencies carefully chosen to produce a bandpass response (Fig. \ref{fig:MBVD}, \ref{fig:BP_f2_Q}a). The series resonators are identical to each other, as are the shunt (parallel) resonators. 

\begin{figure}
    \centering
    \includegraphics[width=8.5cm]{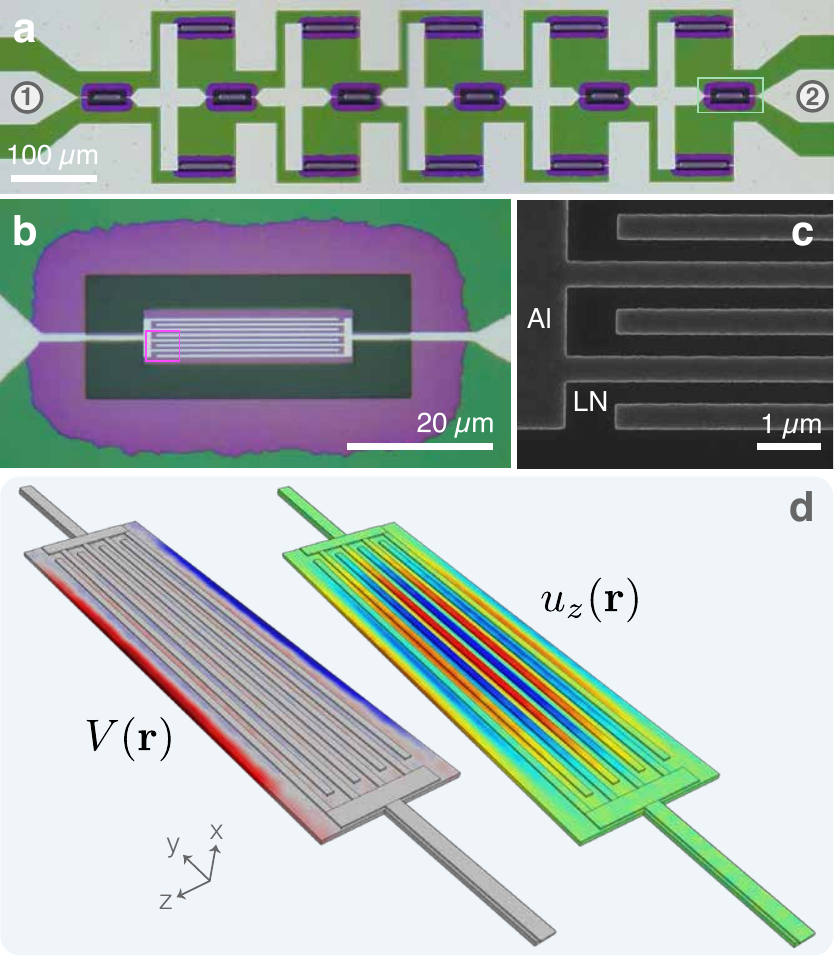}
    \caption{Device layout. \textbf{a.} Optical micrograph of the filter analyzed in Fig. \ref{fig:T1_and_S21}b. Each series resonator (middle row) is shunted by two parallel resonators with total admittance $Y_\text{p}$. This geometry is chosen to symmetrize electromagnetic fields about the signal line. Where the LN film has been released from the silicon substrate, it appears purple, while unreleased regions appear green. Etched gaps in the LN film appear black. Aluminum electrodes and surrounding ground plane (grey) are designed to be contacted by a three-point (ground-signal-ground) coplanar waveguide probe. Transmission measurements in Fig. \ref{fig:T1_and_S21} are made with respect to the indicated ports 1 and 2. \textbf{b.} Optical micrograph detailing the sixth series resonator, highlighted in (a). This resonator has pitch $a = 1.487\,\mu$m, width $W=25\,\mu$m, and $N=4$ IDT pairs. \textbf{c.} Scanning electron micrograph detailing the electrode geometry of (b).  \textbf{d.} Finite-element simulation of a longitudinal Lamb mode at $\omega = 2\pi \times 3.74$ GHz. The z-component of displacement $u_z(\mathbf{r})$ and corresponding electrostatic potential $V(\mathbf{r})$ are plotted. The coordinate system corresponds to the LN crystal axes.}
    \label{fig:fab_pics}
\end{figure}
Each single-resonator electrical response is described by its admittance $Y(\omega)=1/Z(\omega)$ where $Z(\omega)$ is the electrical impedance. This response is well-modeled by a Butterworth-van Dyke (BVD) equivalent circuit (Fig. \ref{fig:MBVD}). \cite{morgan,hashimoto,pop_piazza,lakin} It is important that the antiresonance of the parallel resonators -- the zero in the admittance $Y_\text{p}(\omega)$ -- is placed at the series resonance, or the pole in $Y_\text{s}(\omega)$ (Fig. \ref{fig:BP_f2_Q}a). This frequency defines the center of the passband: here, the parallel resonators have maximal impedance, while the series resonators have minimal impedance, so a microwave signal passes easily through the filter. The spacing between each resonance $f_\text{R}$ and its antiresonance $f_\text{A}$ is given by $k^2 = ({\pi^2}/{8})( {f_\text{A}^2-f_\text{R}^2})/{f_\text{A}^2}$.\cite{pop_piazza} Thus $k^2$, which depends strongly on the material platform, determines filter bandwidth. The BVD circuit elements fully parameterize the frequency ($\omega_\text{m} = {1}/{\sqrt{L\,C}}$), piezoelectric coupling ($ k^2 = {\pi^2}C/{8}{C_\text{g}}$), and quality factor ($Q = {\pi^2}/{8}{\,\omega_\text{m}\,C \,R\,}$).\cite{pop_piazza} %

\begin{figure}
    \centering
    \includegraphics[width=8.5cm]{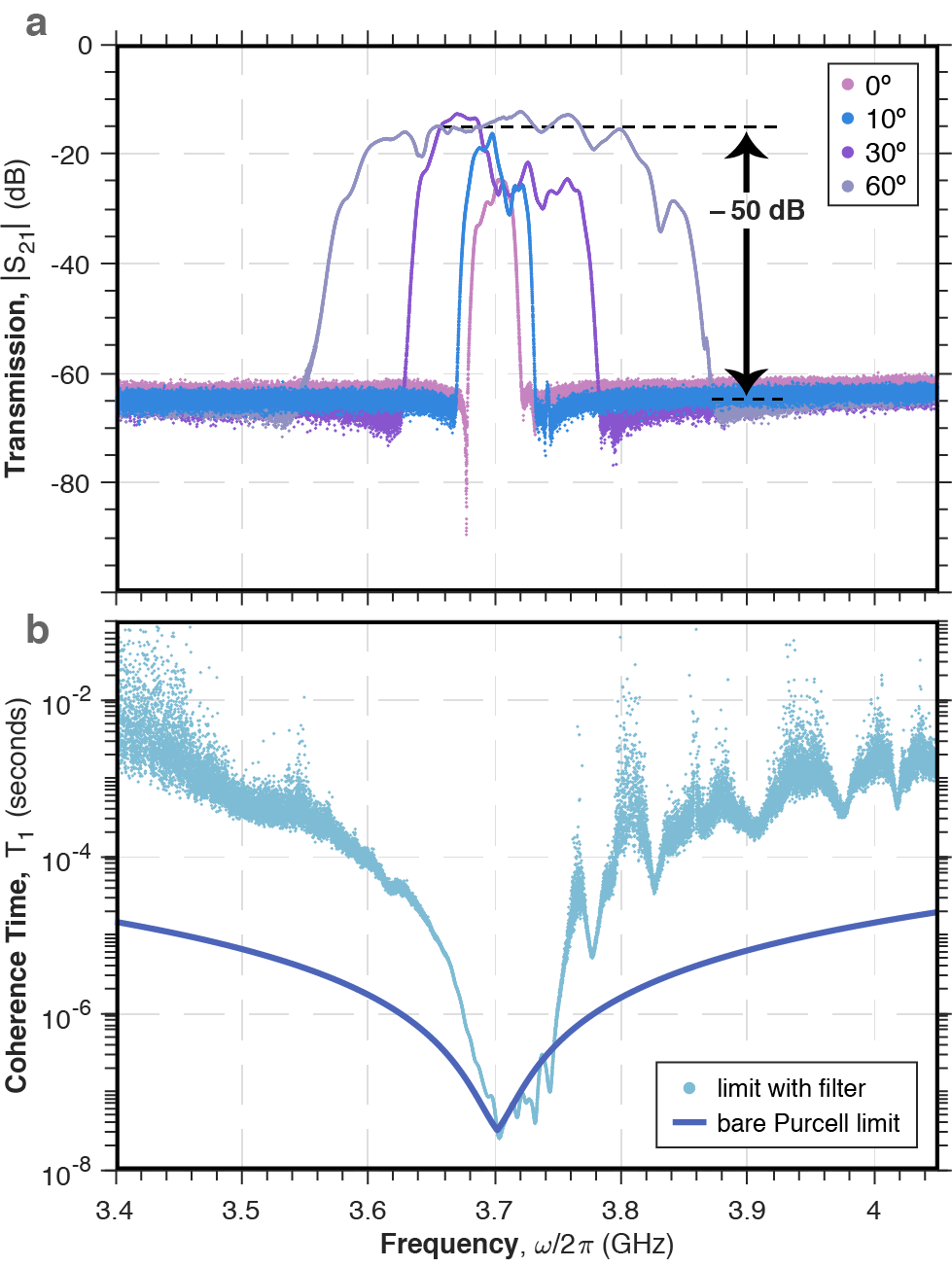}
    \caption{Filter characterization. \textbf{a.} Room temperature transmission spectra for four devices. Each device is rotated (as indicated) counterclockwise to produce an angle $\theta$ between the mechanical propagation direction and the LN extraordinary axis. The center frequencies are shifted by $\pm 100$ MHz to align the passbands for ease of viewing. The labeled x-axis values correspond to the the 10$^\text{o}$ device (blue), which we further analyze with respect to qubit relaxation time. \textbf{b.} Calculated raw Purcell-limited and filter-enhanced coherence times $T_1$ as a function of qubit frequency. This model centers the resonator frequency $\omega_\text{r}= 2\pi \times3.70$ GHz in the filter passband, and assumes constant coupling to the qubit $g= 2\pi \times10$ MHz and resonator decay rate $\kappa = 2\pi \times10$ MHz. Filtered $T_1$ is calculated from Eqs. (\ref{eq:linewidths}) and (\ref{eq:S11_to_Zext}) using $S_{11}$ measurements from the 10$^\text{o}$ device shown in (a).}
    \label{fig:T1_and_S21}
\end{figure}

Intriguingly, we find that while the quality factor of the resonators contributes to insertion loss in the passband as well as less sharply defined band edges, it is not a strong limiting factor in filter performance (Fig. \ref{fig:BP_f2_Q}b). Finally, we note that piezoelectric coupling efficiency can be tuned during fabrication by rotating the orientation of the IDTs with respect to the crystal axes (Fig.~\ref{fig:T1_and_S21}a). It can also be increased by patterning larger transducers, with longer electrodes (wider resonators) or a larger number of interdigitated pairs.

We fabricate our devices on X-cut LN with a process similar to that described in Ref.~\cite{arrangoiz_PRX} We measure the scattering parameters of fabricated filters using a calibrated vector network analyzer (VNA) at room temperature and atmospheric pressure. Reflection and transmission are analyzed to calculate the filter enhancement factor on qubit lifetime.
\begin{center}
    \begin{tabular}{c @{\hspace{.28cm}} c @{\hspace{.28cm}} c @{\hspace{.28cm}} c @{\hspace{.28cm}} c @{\hspace{.28cm}} c @{\hspace{.28cm}} c @{\hspace{.28cm}} c}
         $a_\text{s}$ \footnotesize{($\mu$m)} & $N_\text{s}$ & $W_\text{s}$ \footnotesize{($\mu$m)} & $a_\text{p}$ \footnotesize{($\mu$m)} & $N_\text{p}$ & $W_\text{p}$ \footnotesize{($\mu$m)} & $\theta$ \footnotesize{(º)} & order \\[0.5 ex]
         \hline \hline 
         1.489 & 4 & 15 & 1.500 & 4 & 50 & 0 & 6 \Tstrut \Bstrut\\
        1.487 & 4 & 25 & 1.500 & 4 & 50 & 10 & 6 \Bstrut \\
        1.466 & 4 & 25 & 1.500 & 4 & 50 & 30 & 6 \Bstrut \\
        1.365 & 4 & 25 & 1.500 & 6 & 50 & 60 & 6 \Bstrut \\
    \end{tabular}
\end{center}

$Z_\text{ext}(\omega)$ and consequently the filter factor appearing in Eq. (\ref{eq:linewidths}) can be deduced from calibrated microwave characterization. In particular, the environmental impedance can be extracted from a filter's scattering parameters as:\cite{pozar}
\begin{equation}\label{eq:S11_to_Zext}
\text{Re}\, Z_{\text{ext}} = Z_0\,\frac{1-|S_{11}|^2}{|1-S_{11}|^2}.
\end{equation}
This expression is used to infer the reduction in Purcell decay rate $\gamma_\text{q}$ with the addition of the filter. We calculate unfiltered Purcell-limited as well as filter-enhanced coherence times for a qubit-resonator system with constant $g$ and $\kappa$. The bare Purcell rate is calculated by diagonalizing Eq. (\ref{eq:JC}), using a single-mode resonator model, without assuming $g\ll \Delta$ in the mixing angle $\text{tan}\,2\theta_n = 2g\sqrt n/\Delta$.\cite{schuster_thesis, reed_thesis} We see from Fig.~\ref{fig:T1_and_S21}b that the filter can be used to realize nearly two orders of magnitude enhancement of the qubit lifetime over the unfiltered system. The sharp dips in the filtered $T_1$ spectrum correspond to spurious mechanical resonances of the filter. 

We have proposed and realized mechanical Purcell filters that use nanomechanical elements in a qubit-compatible platform. Using room temperature measurements, we  quantify the expected enhancement of qubit coherence time for a range of qubit frequencies around the center of the passband. We demonstrate that the bandwidths of these filters can be tuned by design, reaching up to 220 MHz, which is broad enough to accommodate many strongly coupled resonators for fast, multiplexed qubit readout. Quantum acoustic systems have been proposed as means of realizing new regimes of quantum optics, quantum memory elements for processors, and quantum state converters for networking. Our work opens a new space in the field by suggesting an application for quantum acoustic systems that can impact development of quantum machines.

The authors thank P.\,Arrangoiz-Arriola and E.\,A.\,Wollack for useful discussions. This work was supported by the U.S. government through the Department of Energy Grant No. DE-SC0019174 and the National Science Foundation Grant No. ECCS-1808100. A.\,Y.\,C. was supported by a Stanford Graduate Fellowship. Part of this work was performed at the Stanford Nano Shared Facilities (SNSF), supported by the National Science Foundation under Grant No. ECCS-1542152, and the Stanford Nanofabrication Facility (SNF).

\nocite{*}
\providecommand{\noopsort}[1]{}\providecommand{\singleletter}[1]{#1
	\providecommand*\hyphen{-} }%


\begin{thebibliography}{33}%
	\makeatletter
	\providecommand \@ifxundefined [1]{%
		\@ifx{#1\undefined}
	}%
	\providecommand \@ifnum [1]{%
		\ifnum #1\expandafter \@firstoftwo
		\else \expandafter \@secondoftwo
		\fi
	}%
	\providecommand \@ifx [1]{%
		\ifx #1\expandafter \@firstoftwo
		\else \expandafter \@secondoftwo
		\fi
	}%
	\providecommand \natexlab [1]{#1}%
	\providecommand \enquote  [1]{``#1''}%
	\providecommand \bibnamefont  [1]{#1}%
	\providecommand \bibfnamefont [1]{#1}%
	\providecommand \citenamefont [1]{#1}%
	\providecommand \href@noop [0]{\@secondoftwo}%
	\providecommand \href [0]{\begingroup \@sanitize@url \@href}%
	\providecommand \@href[1]{\@@startlink{#1}\@@href}%
	\providecommand \@@href[1]{\endgroup#1\@@endlink}%
	\providecommand \@sanitize@url [0]{\catcode `\\12\catcode `\$12\catcode
		`\&12\catcode `\#12\catcode `\^12\catcode `\_12\catcode `\%12\relax}%
	\providecommand \@@startlink[1]{}%
	\providecommand \@@endlink[0]{}%
	\providecommand \url  [0]{\begingroup\@sanitize@url \@url }%
	\providecommand \@url [1]{\endgroup\@href {#1}{\urlprefix }}%
	\providecommand \urlprefix  [0]{URL }%
	\providecommand \Eprint [0]{\href }%
	\providecommand \doibase [0]{http://dx.doi.org/}%
	\providecommand \selectlanguage [0]{\@gobble}%
	\providecommand \bibinfo  [0]{\@secondoftwo}%
	\providecommand \bibfield  [0]{\@secondoftwo}%
	\providecommand \translation [1]{[#1]}%
	\providecommand \BibitemOpen [0]{}%
	\providecommand \bibitemStop [0]{}%
	\providecommand \bibitemNoStop [0]{.\EOS\space}%
	\providecommand \EOS [0]{\spacefactor3000\relax}%
	\providecommand \BibitemShut  [1]{\csname bibitem#1\endcsname}%
	\let\auto@bib@innerbib\@empty
	%
	\bibitem [{\citenamefont {DiVincenzo}(2000)}]{divincenzo}%
	\BibitemOpen
	\bibfield  {author} {\bibinfo {author} {\bibfnamefont {D.~P.}\ \bibnamefont
			{DiVincenzo}},\ }\href@noop {} {\bibfield  {journal} {\bibinfo  {journal}
			{\href{https://arxiv.org/pdf/quant-ph/0002077.pdf}{Fortschritte der Physik}}\
		}\textbf {\bibinfo {volume} {48}},\ \bibinfo {pages} {9} (\bibinfo {year}
		{2000})}\BibitemShut {NoStop}%
	\bibitem [{\citenamefont {Devoret}\ and\ \citenamefont
		{Schoelkopf}(2013)}]{devoret_schoelkopf}%
	\BibitemOpen
	\bibfield  {author} {\bibinfo {author} {\bibfnamefont {M.~H.}\ \bibnamefont
			{Devoret}}\ and\ \bibinfo {author} {\bibfnamefont {R.~J.}\ \bibnamefont
			{Schoelkopf}},\ }\href@noop {} {\bibfield  {journal} {\bibinfo  {journal}
			{\href{https://science.sciencemag.org/content/339/6124/1169}{Science}}\
		}\textbf {\bibinfo {volume} {339}},\ \bibinfo {pages} {1169} (\bibinfo {year}
		{2013})}\BibitemShut {NoStop}%
	\bibitem [{\citenamefont {Koch}\ \emph {et~al.}(2007)\citenamefont {Koch},
		\citenamefont {Yu}, \citenamefont {Gambetta}, \citenamefont {Houck},
		\citenamefont {Schuster}, \citenamefont {Majer}, \citenamefont {Blais},
		\citenamefont {Devoret}, \citenamefont {Girvin},\ and\ \citenamefont
		{Schoelkopf}}]{koch_schoelkopf}%
	\BibitemOpen
	\bibfield  {author} {\bibinfo {author} {\bibfnamefont {J.}~\bibnamefont
			{Koch}}, \bibinfo {author} {\bibfnamefont {T.~M.}\ \bibnamefont {Yu}},
		\bibinfo {author} {\bibfnamefont {J.}~\bibnamefont {Gambetta}}, \bibinfo
		{author} {\bibfnamefont {A.~A.}\ \bibnamefont {Houck}}, \bibinfo {author}
		{\bibfnamefont {D.~I.}\ \bibnamefont {Schuster}}, \bibinfo {author}
		{\bibfnamefont {J.}~\bibnamefont {Majer}}, \bibinfo {author} {\bibfnamefont
			{A.}~\bibnamefont {Blais}}, \bibinfo {author} {\bibfnamefont {M.~H.}\
			\bibnamefont {Devoret}}, \bibinfo {author} {\bibfnamefont {S.~M.}\
			\bibnamefont {Girvin}}, \ and\ \bibinfo {author} {\bibfnamefont {R.~J.}\
			\bibnamefont {Schoelkopf}},\ }\href@noop {} {\bibfield  {journal} {\bibinfo
			{journal}
			{\href{https://journals.aps.org/pra/pdf/10.1103/PhysRevA.76.042319}{Phys.
					Rev. A}}\ }\textbf {\bibinfo {volume} {76}},\ \bibinfo {pages} {042319}
		(\bibinfo {year} {2007})}\BibitemShut {NoStop}%
	\bibitem [{\citenamefont {Neill}\ \emph {et~al.}(2018)\citenamefont {Neill},
		\citenamefont {Roushan}, \citenamefont {Kechedzhi}, \citenamefont {Boixo},
		\citenamefont {Isakov}, \citenamefont {Smelyanskiy}, \citenamefont {Megrant},
		\citenamefont {Chiaro}, \citenamefont {Dunsworth}, \citenamefont {Arya},
		\citenamefont {Barends}, \citenamefont {Burkett}, \citenamefont {Chen},
		\citenamefont {Chen}, \citenamefont {Fowler}, \citenamefont {Foxen},
		\citenamefont {Giustina}, \citenamefont {Graff}, \citenamefont {Jeffrey},
		\citenamefont {Huang}, \citenamefont {Kelly}, \citenamefont {Klimov},
		\citenamefont {Lucero}, \citenamefont {Mutus}, \citenamefont {Neeley},
		\citenamefont {Quintana}, \citenamefont {Sank}, \citenamefont {Vainsencher},
		\citenamefont {Wenner}, \citenamefont {White}, \citenamefont {Neven},\ and\
		\citenamefont {Martinis}}]{neill_martinis}%
	\BibitemOpen
	\bibfield  {author} {\bibinfo {author} {\bibfnamefont {C.}~\bibnamefont
			{Neill}}, \bibinfo {author} {\bibfnamefont {P.}~\bibnamefont {Roushan}},
		\bibinfo {author} {\bibfnamefont {K.}~\bibnamefont {Kechedzhi}}, \bibinfo
		{author} {\bibfnamefont {S.}~\bibnamefont {Boixo}}, \bibinfo {author}
		{\bibfnamefont {S.~V.}\ \bibnamefont {Isakov}}, \bibinfo {author}
		{\bibfnamefont {V.}~\bibnamefont {Smelyanskiy}}, \bibinfo {author}
		{\bibfnamefont {A.}~\bibnamefont {Megrant}}, \bibinfo {author} {\bibfnamefont
			{B.}~\bibnamefont {Chiaro}}, \bibinfo {author} {\bibfnamefont
			{A.}~\bibnamefont {Dunsworth}}, \bibinfo {author} {\bibfnamefont
			{K.}~\bibnamefont {Arya}}, \bibinfo {author} {\bibfnamefont {R.}~\bibnamefont
			{Barends}}, \bibinfo {author} {\bibfnamefont {B.}~\bibnamefont {Burkett}},
		\bibinfo {author} {\bibfnamefont {Y.}~\bibnamefont {Chen}}, \bibinfo {author}
		{\bibfnamefont {Z.}~\bibnamefont {Chen}}, \bibinfo {author} {\bibfnamefont
			{A.}~\bibnamefont {Fowler}}, \bibinfo {author} {\bibfnamefont
			{B.}~\bibnamefont {Foxen}}, \bibinfo {author} {\bibfnamefont
			{M.}~\bibnamefont {Giustina}}, \bibinfo {author} {\bibfnamefont
			{R.}~\bibnamefont {Graff}}, \bibinfo {author} {\bibfnamefont
			{E.}~\bibnamefont {Jeffrey}}, \bibinfo {author} {\bibfnamefont
			{T.}~\bibnamefont {Huang}}, \bibinfo {author} {\bibfnamefont
			{J.}~\bibnamefont {Kelly}}, \bibinfo {author} {\bibfnamefont
			{P.}~\bibnamefont {Klimov}}, \bibinfo {author} {\bibfnamefont
			{E.}~\bibnamefont {Lucero}}, \bibinfo {author} {\bibfnamefont
			{J.}~\bibnamefont {Mutus}}, \bibinfo {author} {\bibfnamefont
			{M.}~\bibnamefont {Neeley}}, \bibinfo {author} {\bibfnamefont
			{C.}~\bibnamefont {Quintana}}, \bibinfo {author} {\bibfnamefont
			{D.}~\bibnamefont {Sank}}, \bibinfo {author} {\bibfnamefont {A.}~\bibnamefont
			{Vainsencher}}, \bibinfo {author} {\bibfnamefont {J.}~\bibnamefont {Wenner}},
		\bibinfo {author} {\bibfnamefont {T.~C.}\ \bibnamefont {White}}, \bibinfo
		{author} {\bibfnamefont {H.}~\bibnamefont {Neven}}, \ and\ \bibinfo {author}
		{\bibfnamefont {J.~M.}\ \bibnamefont {Martinis}},\ }\href@noop {} {\bibfield
		{journal} {\bibinfo  {journal}
			{\href{https://science.sciencemag.org/content/360/6385/195/tab-pdf}{Science}}\
		}\textbf {\bibinfo {volume} {360}},\ \bibinfo {pages} {195} (\bibinfo {year}
		{2018})}\BibitemShut {NoStop}%
	\bibitem [{\citenamefont {Purcell}(1946)}]{purcell}%
	\BibitemOpen
	\bibfield  {author} {\bibinfo {author} {\bibfnamefont {E.~M.}\ \bibnamefont
			{Purcell}},\ }\href@noop {} {\bibfield  {journal} {\bibinfo  {journal}
			{\href{https://link.springer.com/content/pdf/10.1007/978-1-4615-1963-8_40.pdf}{Phys.\
					Rev.}}\ }\textbf {\bibinfo {volume} {69}},\ \bibinfo {pages} {681} (\bibinfo
		{year} {1946})}\BibitemShut {NoStop}%
	\bibitem [{\citenamefont {Goy}\ \emph {et~al.}(1983)\citenamefont {Goy},
		\citenamefont {Haimond}, \citenamefont {Gross},\ and\ \citenamefont
		{Haroche}}]{goy_haroche}%
	\BibitemOpen
	\bibfield  {author} {\bibinfo {author} {\bibfnamefont {P.}~\bibnamefont
			{Goy}}, \bibinfo {author} {\bibfnamefont {J.~M.}\ \bibnamefont {Haimond}},
		\bibinfo {author} {\bibfnamefont {M.}~\bibnamefont {Gross}}, \ and\ \bibinfo
		{author} {\bibfnamefont {S.}~\bibnamefont {Haroche}},\ }\href@noop {}
	{\bibfield  {journal} {\bibinfo  {journal}
			{\href{https://journals.aps.org/prl/pdf/10.1103/PhysRevLett.50.1903}{Phys.
					Rev. Lett.}}\ }\textbf {\bibinfo {volume} {50}},\ \bibinfo {pages} {1903}
		(\bibinfo {year} {1983})}\BibitemShut {NoStop}%
	\bibitem [{\citenamefont {Girvin}(2014)}]{girvin}%
	\BibitemOpen
	\bibfield  {author} {\bibinfo {author} {\bibfnamefont {S.~M.}\ \bibnamefont
			{Girvin}},\ }\bibfield  {title} {\enquote {\bibinfo {title} {Circuit {QED}:
				Superconducting qubits coupled to microwave photons},}\ }in\ \href@noop {}
	{\emph {\bibinfo {booktitle} {Quantum Machines: Measurement and Control of
				Engineered Quantum Systems}}},\ \bibinfo {editor} {edited by\ \bibinfo
		{editor} {\bibfnamefont {M.}~\bibnamefont {Devoret}}, \bibinfo {editor}
		{\bibfnamefont {B.}~\bibnamefont {Huard}}, \bibinfo {editor} {\bibfnamefont
			{R.}~\bibnamefont {Schoelkopf}}, \ and\ \bibinfo {editor} {\bibfnamefont
			{L.}~\bibnamefont {Cugliandolo}}}\ (\bibinfo  {publisher} {Oxford University
		Press},\ \bibinfo {year} {2014})\BibitemShut {NoStop}%
	\bibitem [{\citenamefont {Didier}, \citenamefont {Bourassa},\ and\
		\citenamefont {Blais}(2015)}]{non_dispersive}%
	\BibitemOpen
	\bibfield  {author} {\bibinfo {author} {\bibfnamefont {N.}~\bibnamefont
			{Didier}}, \bibinfo {author} {\bibfnamefont {J.}~\bibnamefont {Bourassa}}, \
		and\ \bibinfo {author} {\bibfnamefont {A.}~\bibnamefont {Blais}},\
	}\href@noop {} {\bibfield  {journal} {\bibinfo  {journal}
			{\href{https://journals.aps.org/prl/pdf/10.1103/PhysRevLett.115.203601}{Phys.
					Rev. Lett.}}\ }\textbf {\bibinfo {volume} {115}},\ \bibinfo {pages} {203601}
		(\bibinfo {year} {2015})}\BibitemShut {NoStop}%
	\bibitem [{\citenamefont {Vijay}, \citenamefont {Slichter},\ and\ \citenamefont
		{Siddiqi}(2011)}]{quantum_jumps_siddiqi}%
	\BibitemOpen
	\bibfield  {author} {\bibinfo {author} {\bibfnamefont {R.}~\bibnamefont
			{Vijay}}, \bibinfo {author} {\bibfnamefont {D.~H.}\ \bibnamefont {Slichter}},
		\ and\ \bibinfo {author} {\bibfnamefont {I.}~\bibnamefont {Siddiqi}},\
	}\href@noop {} {\bibfield  {journal} {\bibinfo  {journal}
			{\href{https://journals.aps.org/prl/pdf/10.1103/PhysRevLett.106.110502}{Phys.\,Rev.\,Lett.}}\
		}\textbf {\bibinfo {volume} {106}},\ \bibinfo {pages} {110502} (\bibinfo
		{year} {2011})}\BibitemShut {NoStop}%
	\bibitem [{\citenamefont {Johnson}\ \emph {et~al.}(2012)\citenamefont
		{Johnson}, \citenamefont {Macklin}, \citenamefont {Slichter}, \citenamefont
		{Vijay}, \citenamefont {Weingarten}, \citenamefont {Clarke},\ and\
		\citenamefont {Siddiqi}}]{johnson_clarke_siddiqi}%
	\BibitemOpen
	\bibfield  {author} {\bibinfo {author} {\bibfnamefont {J.~E.}\ \bibnamefont
			{Johnson}}, \bibinfo {author} {\bibfnamefont {C.}~\bibnamefont {Macklin}},
		\bibinfo {author} {\bibfnamefont {D.~H.}\ \bibnamefont {Slichter}}, \bibinfo
		{author} {\bibfnamefont {R.}~\bibnamefont {Vijay}}, \bibinfo {author}
		{\bibfnamefont {E.~B.}\ \bibnamefont {Weingarten}}, \bibinfo {author}
		{\bibfnamefont {J.}~\bibnamefont {Clarke}}, \ and\ \bibinfo {author}
		{\bibfnamefont {I.}~\bibnamefont {Siddiqi}},\ }\href@noop {} {\bibfield
		{journal} {\bibinfo  {journal}
			{\href{https://journals.aps.org/prl/pdf/10.1103/PhysRevLett.109.050506}{Phys.\,Rev.\,Lett.}}\
		}\textbf {\bibinfo {volume} {109}},\ \bibinfo {pages} {050506} (\bibinfo
		{year} {2012})}\BibitemShut {NoStop}%
	\bibitem [{\citenamefont {Houck}\ \emph {et~al.}(2007)\citenamefont {Houck},
		\citenamefont {Schuster}, \citenamefont {Gambetta}, \citenamefont {Schreier},
		\citenamefont {Johnson}, \citenamefont {Chow}, \citenamefont {Frunzio},
		\citenamefont {Majer}, \citenamefont {Devoret}, \citenamefont {Girvin},\ and\
		\citenamefont {Schoelkopf}}]{houck_single_photons}%
	\BibitemOpen
	\bibfield  {author} {\bibinfo {author} {\bibfnamefont {A.~A.}\ \bibnamefont
			{Houck}}, \bibinfo {author} {\bibfnamefont {D.~I.}\ \bibnamefont {Schuster}},
		\bibinfo {author} {\bibfnamefont {J.~M.}\ \bibnamefont {Gambetta}}, \bibinfo
		{author} {\bibfnamefont {J.~A.}\ \bibnamefont {Schreier}}, \bibinfo {author}
		{\bibfnamefont {B.~R.}\ \bibnamefont {Johnson}}, \bibinfo {author}
		{\bibfnamefont {J.~M.}\ \bibnamefont {Chow}}, \bibinfo {author}
		{\bibfnamefont {L.}~\bibnamefont {Frunzio}}, \bibinfo {author} {\bibfnamefont
			{J.}~\bibnamefont {Majer}}, \bibinfo {author} {\bibfnamefont {M.~H.}\
			\bibnamefont {Devoret}}, \bibinfo {author} {\bibfnamefont {S.~M.}\
			\bibnamefont {Girvin}}, \ and\ \bibinfo {author} {\bibfnamefont {R.~J.}\
			\bibnamefont {Schoelkopf}},\ }\href@noop {} {\bibfield  {journal} {\bibinfo
			{journal} {\href{https://www.nature.com/articles/nature06126.pdf}{Nature}}\
		}\textbf {\bibinfo {volume} {449}},\ \bibinfo {pages} {328} (\bibinfo {year}
		{2007})}\BibitemShut {NoStop}%
	\bibitem [{\citenamefont {Gambetta}, \citenamefont {Houck},\ and\ \citenamefont
		{Blais}(2011)}]{gambetta_blais}%
	\BibitemOpen
	\bibfield  {author} {\bibinfo {author} {\bibfnamefont {J.~M.}\ \bibnamefont
			{Gambetta}}, \bibinfo {author} {\bibfnamefont {A.~A.}\ \bibnamefont {Houck}},
		\ and\ \bibinfo {author} {\bibfnamefont {A.}~\bibnamefont {Blais}},\
	}\href@noop {} {\bibfield  {journal} {\bibinfo  {journal}
			{\href{https://journals.aps.org/prl/pdf/10.1103/PhysRevLett.106.030502}{Phys.
					Rev. Lett.}}\ }\textbf {\bibinfo {volume} {106}},\ \bibinfo {pages} {030502}
		(\bibinfo {year} {2011})}\BibitemShut {NoStop}%
	\bibitem [{\citenamefont {Houck}\ \emph {et~al.}(2008)\citenamefont {Houck},
		\citenamefont {Schrier}, \citenamefont {Johnson}, \citenamefont {Chow},
		\citenamefont {Koch}, \citenamefont {Gambetta}, \citenamefont {Schuster},
		\citenamefont {Frunzio}, \citenamefont {Devoret}, \citenamefont {Girvin},\
		and\ \citenamefont {Schoelkopf}}]{houck_schoelkopf}%
	\BibitemOpen
	\bibfield  {author} {\bibinfo {author} {\bibfnamefont {A.~A.}\ \bibnamefont
			{Houck}}, \bibinfo {author} {\bibfnamefont {J.~A.}\ \bibnamefont {Schrier}},
		\bibinfo {author} {\bibfnamefont {B.~R.}\ \bibnamefont {Johnson}}, \bibinfo
		{author} {\bibfnamefont {J.~M.}\ \bibnamefont {Chow}}, \bibinfo {author}
		{\bibfnamefont {J.}~\bibnamefont {Koch}}, \bibinfo {author} {\bibfnamefont
			{J.~M.}\ \bibnamefont {Gambetta}}, \bibinfo {author} {\bibfnamefont {D.~I.}\
			\bibnamefont {Schuster}}, \bibinfo {author} {\bibfnamefont {L.}~\bibnamefont
			{Frunzio}}, \bibinfo {author} {\bibfnamefont {M.~H.}\ \bibnamefont
			{Devoret}}, \bibinfo {author} {\bibfnamefont {S.~M.}\ \bibnamefont {Girvin}},
		\ and\ \bibinfo {author} {\bibfnamefont {R.~J.}\ \bibnamefont {Schoelkopf}},\
	}\href@noop {} {\bibfield  {journal} {\bibinfo  {journal}
			{\href{https://journals.aps.org/prl/pdf/10.1103/PhysRevLett.101.080502}{Phys.\
					Rev.\ Lett.}}\ }\textbf {\bibinfo {volume} {101}},\ \bibinfo {pages} {080502}
		(\bibinfo {year} {2008})}\BibitemShut {NoStop}%
	\bibitem [{\citenamefont {Reed}\ \emph {et~al.}(2010)\citenamefont {Reed},
		\citenamefont {Johnson}, \citenamefont {Houck}, \citenamefont {DiCarlo},
		\citenamefont {Chow}, \citenamefont {Schuster}, \citenamefont {Frunzio},\
		and\ \citenamefont {Schoelkopf}}]{reed_schoelkopf}%
	\BibitemOpen
	\bibfield  {author} {\bibinfo {author} {\bibfnamefont {M.~D.}\ \bibnamefont
			{Reed}}, \bibinfo {author} {\bibfnamefont {B.~R.}\ \bibnamefont {Johnson}},
		\bibinfo {author} {\bibfnamefont {A.~A.}\ \bibnamefont {Houck}}, \bibinfo
		{author} {\bibfnamefont {L.}~\bibnamefont {DiCarlo}}, \bibinfo {author}
		{\bibfnamefont {J.~M.}\ \bibnamefont {Chow}}, \bibinfo {author}
		{\bibfnamefont {D.~I.}\ \bibnamefont {Schuster}}, \bibinfo {author}
		{\bibfnamefont {L.}~\bibnamefont {Frunzio}}, \ and\ \bibinfo {author}
		{\bibfnamefont {R.~J.}\ \bibnamefont {Schoelkopf}},\ }\href@noop {}
	{\bibfield  {journal} {\bibinfo  {journal}
			{\href{https://aip.scitation.org/doi/pdf/10.1063/1.3435463?class=pdf}{Appl.\
					Phys.\ Lett.}}\ }\textbf {\bibinfo {volume} {96}},\ \bibinfo {pages} {203110}
		(\bibinfo {year} {2010})}\BibitemShut {NoStop}%
	\bibitem [{\citenamefont {Jeffrey}\ \emph {et~al.}(2014)\citenamefont
		{Jeffrey}, \citenamefont {Sank}, \citenamefont {Mutus}, \citenamefont
		{White}, \citenamefont {Kelly}, \citenamefont {Barends}, \citenamefont
		{Chen}, \citenamefont {Chen}, \citenamefont {Chiaro}, \citenamefont
		{Dunsworth}, \citenamefont {Megrant}, \citenamefont {O'Malley}, \citenamefont
		{Neill}, \citenamefont {Roushan}, \citenamefont {Vainsencher}, \citenamefont
		{Wenner}, \citenamefont {Cleland},\ and\ \citenamefont
		{Martinis}}]{jeffrey_martinis}%
	\BibitemOpen
	\bibfield  {author} {\bibinfo {author} {\bibfnamefont {E.}~\bibnamefont
			{Jeffrey}}, \bibinfo {author} {\bibfnamefont {D.}~\bibnamefont {Sank}},
		\bibinfo {author} {\bibfnamefont {J.~Y.}\ \bibnamefont {Mutus}}, \bibinfo
		{author} {\bibfnamefont {T.~C.}\ \bibnamefont {White}}, \bibinfo {author}
		{\bibfnamefont {J.}~\bibnamefont {Kelly}}, \bibinfo {author} {\bibfnamefont
			{R.}~\bibnamefont {Barends}}, \bibinfo {author} {\bibfnamefont
			{Y.}~\bibnamefont {Chen}}, \bibinfo {author} {\bibfnamefont {Z.}~\bibnamefont
			{Chen}}, \bibinfo {author} {\bibfnamefont {B.}~\bibnamefont {Chiaro}},
		\bibinfo {author} {\bibfnamefont {A.}~\bibnamefont {Dunsworth}}, \bibinfo
		{author} {\bibfnamefont {A.}~\bibnamefont {Megrant}}, \bibinfo {author}
		{\bibfnamefont {P.~J.~J.}\ \bibnamefont {O'Malley}}, \bibinfo {author}
		{\bibfnamefont {C.}~\bibnamefont {Neill}}, \bibinfo {author} {\bibfnamefont
			{P.}~\bibnamefont {Roushan}}, \bibinfo {author} {\bibfnamefont
			{A.}~\bibnamefont {Vainsencher}}, \bibinfo {author} {\bibfnamefont
			{J.}~\bibnamefont {Wenner}}, \bibinfo {author} {\bibfnamefont {A.~N.}\
			\bibnamefont {Cleland}}, \ and\ \bibinfo {author} {\bibfnamefont {J.~M.}\
			\bibnamefont {Martinis}},\ }\href@noop {} {\bibfield  {journal} {\bibinfo
			{journal}
			{\href{https://journals.aps.org/prl/pdf/10.1103/PhysRevLett.112.190504}{Phys.\
					Rev.\ Lett.}}\ }\textbf {\bibinfo {volume} {112}},\ \bibinfo {pages} {190504}
		(\bibinfo {year} {2014})}\BibitemShut {NoStop}%
	\bibitem [{\citenamefont {Sete}, \citenamefont {Martinis},\ and\ \citenamefont
		{Korotkov}(2015)}]{sete_martinis}%
	\BibitemOpen
	\bibfield  {author} {\bibinfo {author} {\bibfnamefont {E.~A.}\ \bibnamefont
			{Sete}}, \bibinfo {author} {\bibfnamefont {J.~M.}\ \bibnamefont {Martinis}},
		\ and\ \bibinfo {author} {\bibfnamefont {A.~N.}\ \bibnamefont {Korotkov}},\
	}\href@noop {} {\bibfield  {journal} {\bibinfo  {journal}
			{\href{https://journals.aps.org/pra/pdf/10.1103/PhysRevA.92.012325}{Phys.\
					Rev.\ A}}\ }\textbf {\bibinfo {volume} {92}},\ \bibinfo {pages} {012325}
		(\bibinfo {year} {2015})}\BibitemShut {NoStop}%
	\bibitem [{\citenamefont {Bronn}\ \emph {et~al.}(2015)\citenamefont {Bronn},
		\citenamefont {Liu}, \citenamefont {Hertzberg}, \citenamefont {C{\'o}rcoles},
		\citenamefont {Houck}, \citenamefont {Gambetta},\ and\ \citenamefont
		{Chow}}]{bronn_chow}%
	\BibitemOpen
	\bibfield  {author} {\bibinfo {author} {\bibfnamefont {N.~T.}\ \bibnamefont
			{Bronn}}, \bibinfo {author} {\bibfnamefont {Y.}~\bibnamefont {Liu}}, \bibinfo
		{author} {\bibfnamefont {J.~B.}\ \bibnamefont {Hertzberg}}, \bibinfo {author}
		{\bibfnamefont {A.~D.}\ \bibnamefont {C{\'o}rcoles}}, \bibinfo {author}
		{\bibfnamefont {A.~A.}\ \bibnamefont {Houck}}, \bibinfo {author}
		{\bibfnamefont {J.~M.}\ \bibnamefont {Gambetta}}, \ and\ \bibinfo {author}
		{\bibfnamefont {J.~M.}\ \bibnamefont {Chow}},\ }\href@noop {} {\bibfield
		{journal} {\bibinfo  {journal}
			{\href{https://aip.scitation.org/doi/pdf/10.1063/1.4934867?class=pdf}{Appl.\
					Phys.\ Lett.}}\ }\textbf {\bibinfo {volume} {107}},\ \bibinfo {pages}
		{172601} (\bibinfo {year} {2015})}\BibitemShut {NoStop}%
	\bibitem [{\citenamefont {Walter}\ \emph {et~al.}(2017)\citenamefont {Walter},
		\citenamefont {Kurpiers}, \citenamefont {Gasparinetti}, \citenamefont
		{Magnard}, \citenamefont {Poto{\u c}nik}, \citenamefont {Salath{\' e}},
		\citenamefont {Pechal}, \citenamefont {Mondal}, \citenamefont {Oppliger},
		\citenamefont {Eichler},\ and\ \citenamefont {Wallraff}}]{walter_wallraff}%
	\BibitemOpen
	\bibfield  {author} {\bibinfo {author} {\bibfnamefont {T.}~\bibnamefont
			{Walter}}, \bibinfo {author} {\bibfnamefont {P.}~\bibnamefont {Kurpiers}},
		\bibinfo {author} {\bibfnamefont {S.}~\bibnamefont {Gasparinetti}}, \bibinfo
		{author} {\bibfnamefont {P.}~\bibnamefont {Magnard}}, \bibinfo {author}
		{\bibfnamefont {A.}~\bibnamefont {Poto{\u c}nik}}, \bibinfo {author}
		{\bibfnamefont {Y.}~\bibnamefont {Salath{\' e}}}, \bibinfo {author}
		{\bibfnamefont {M.}~\bibnamefont {Pechal}}, \bibinfo {author} {\bibfnamefont
			{M.}~\bibnamefont {Mondal}}, \bibinfo {author} {\bibfnamefont
			{M.}~\bibnamefont {Oppliger}}, \bibinfo {author} {\bibfnamefont
			{C.}~\bibnamefont {Eichler}}, \ and\ \bibinfo {author} {\bibfnamefont
			{A.}~\bibnamefont {Wallraff}},\ }\href@noop {} {\bibfield  {journal}
		{\bibinfo  {journal}
			{\href{https://journals.aps.org/prapplied/pdf/10.1103/PhysRevApplied.7.054020}{Phys.\
					Rev.\ Appl.}}\ }\textbf {\bibinfo {volume} {7}},\ \bibinfo {pages} {054020}
		(\bibinfo {year} {2017})}\BibitemShut {NoStop}%
	\bibitem [{\citenamefont {Heinsoo}\ \emph {et~al.}(2018)\citenamefont
		{Heinsoo}, \citenamefont {Andersen}, \citenamefont {Remm}, \citenamefont
		{Krinner}, \citenamefont {Walter}, \citenamefont {Salathé}, \citenamefont
		{Gasparinetti}, \citenamefont {Besse}, \citenamefont {Poto{\u c}nik},
		\citenamefont {Wallraff},\ and\ \citenamefont {Eichler}}]{heinsoo_wallraff}%
	\BibitemOpen
	\bibfield  {author} {\bibinfo {author} {\bibfnamefont {J.}~\bibnamefont
			{Heinsoo}}, \bibinfo {author} {\bibfnamefont {C.~K.}\ \bibnamefont
			{Andersen}}, \bibinfo {author} {\bibfnamefont {A.}~\bibnamefont {Remm}},
		\bibinfo {author} {\bibfnamefont {S.}~\bibnamefont {Krinner}}, \bibinfo
		{author} {\bibfnamefont {T.}~\bibnamefont {Walter}}, \bibinfo {author}
		{\bibfnamefont {Y.}~\bibnamefont {Salathé}}, \bibinfo {author}
		{\bibfnamefont {S.}~\bibnamefont {Gasparinetti}}, \bibinfo {author}
		{\bibfnamefont {J.-C.}\ \bibnamefont {Besse}}, \bibinfo {author}
		{\bibfnamefont {A.}~\bibnamefont {Poto{\u c}nik}}, \bibinfo {author}
		{\bibfnamefont {A.}~\bibnamefont {Wallraff}}, \ and\ \bibinfo {author}
		{\bibfnamefont {C.}~\bibnamefont {Eichler}},\ }\href@noop {} {\bibfield
		{journal} {\bibinfo  {journal}
			{\href{https://journals.aps.org/prapplied/pdf/10.1103/PhysRevApplied.10.034040}{Phys.\
					Rev.\ Appl.}}\ }\textbf {\bibinfo {volume} {10}},\ \bibinfo {pages} {034040}
		(\bibinfo {year} {2018})}\BibitemShut {NoStop}%
	\bibitem [{\citenamefont {Arrangoiz-Arriola}\ and\ \citenamefont
		{Safavi-Naeini}(2016)}]{arrangoiz_PRA}%
	\BibitemOpen
	\bibfield  {author} {\bibinfo {author} {\bibfnamefont {P.}~\bibnamefont
			{Arrangoiz-Arriola}}\ and\ \bibinfo {author} {\bibfnamefont {A.~H.}\
			\bibnamefont {Safavi-Naeini}},\ }\href@noop {} {\bibfield  {journal}
		{\bibinfo  {journal}
			{\href{https://journals.aps.org/pra/pdf/10.1103/PhysRevA.94.063864}{Phys.\,Rev.\,A}}\
		}\textbf {\bibinfo {volume} {94}},\ \bibinfo {pages} {063864} (\bibinfo
		{year} {2016})}\BibitemShut {NoStop}%
	\bibitem [{\citenamefont {Satzinger}\ \emph {et~al.}(2018)\citenamefont
		{Satzinger}, \citenamefont {Zhong}, \citenamefont {Chang}, \citenamefont
		{Peairs}, \citenamefont {Bienfait}, \citenamefont {Chou}, \citenamefont
		{Cleland}, \citenamefont {Conner}, \citenamefont {Dumur}, \citenamefont
		{Grebel}, \citenamefont {Gutierrez}, \citenamefont {November}, \citenamefont
		{Povey}, \citenamefont {Whiteley}, \citenamefont {Awschalom}, \citenamefont
		{Schuster},\ and\ \citenamefont {Cleland}}]{satzinger_cleland}%
	\BibitemOpen
	\bibfield  {author} {\bibinfo {author} {\bibfnamefont {K.~J.}\ \bibnamefont
			{Satzinger}}, \bibinfo {author} {\bibfnamefont {Y.~P.}\ \bibnamefont
			{Zhong}}, \bibinfo {author} {\bibfnamefont {H.-S.}\ \bibnamefont {Chang}},
		\bibinfo {author} {\bibfnamefont {G.~A.}\ \bibnamefont {Peairs}}, \bibinfo
		{author} {\bibfnamefont {A.}~\bibnamefont {Bienfait}}, \bibinfo {author}
		{\bibfnamefont {M.-H.}\ \bibnamefont {Chou}}, \bibinfo {author}
		{\bibfnamefont {A.~Y.}\ \bibnamefont {Cleland}}, \bibinfo {author}
		{\bibfnamefont {C.~R.}\ \bibnamefont {Conner}}, \bibinfo {author}
		{\bibfnamefont {{\'E}.}~\bibnamefont {Dumur}}, \bibinfo {author}
		{\bibfnamefont {J.}~\bibnamefont {Grebel}}, \bibinfo {author} {\bibfnamefont
			{I.}~\bibnamefont {Gutierrez}}, \bibinfo {author} {\bibfnamefont {B.~H.}\
			\bibnamefont {November}}, \bibinfo {author} {\bibfnamefont {R.~G.}\
			\bibnamefont {Povey}}, \bibinfo {author} {\bibfnamefont {S.~J.}\ \bibnamefont
			{Whiteley}}, \bibinfo {author} {\bibfnamefont {D.~D.}\ \bibnamefont
			{Awschalom}}, \bibinfo {author} {\bibfnamefont {D.~I.}\ \bibnamefont
			{Schuster}}, \ and\ \bibinfo {author} {\bibfnamefont {A.~N.}\ \bibnamefont
			{Cleland}},\ }\href@noop {} {\bibfield  {journal} {\bibinfo  {journal}
			{\href{https://www.nature.com/articles/s41586-018-0719-5.pdf}{Nature}}\
		}\textbf {\bibinfo {volume} {563}},\ \bibinfo {pages} {661} (\bibinfo {year}
		{2018})}\BibitemShut {NoStop}%
	\bibitem [{\citenamefont {Chu}\ \emph {et~al.}(2017)\citenamefont {Chu},
		\citenamefont {Kharel}, \citenamefont {Renninger}, \citenamefont {Burkhart},
		\citenamefont {Frunzio}, \citenamefont {Rakich},\ and\ \citenamefont
		{Schoelkopf}}]{chu_schoelkopf}%
	\BibitemOpen
	\bibfield  {author} {\bibinfo {author} {\bibfnamefont {Y.}~\bibnamefont
			{Chu}}, \bibinfo {author} {\bibfnamefont {P.}~\bibnamefont {Kharel}},
		\bibinfo {author} {\bibfnamefont {W.~H.}\ \bibnamefont {Renninger}}, \bibinfo
		{author} {\bibfnamefont {L.~D.}\ \bibnamefont {Burkhart}}, \bibinfo {author}
		{\bibfnamefont {L.}~\bibnamefont {Frunzio}}, \bibinfo {author} {\bibfnamefont
			{P.~T.}\ \bibnamefont {Rakich}}, \ and\ \bibinfo {author} {\bibfnamefont
			{R.~J.}\ \bibnamefont {Schoelkopf}},\ }\href@noop {} {\bibfield  {journal}
		{\bibinfo  {journal}
			{\href{https://science.sciencemag.org/content/sci/358/6360/199.full.pdf}{Science}}\
		}\textbf {\bibinfo {volume} {13}},\ \bibinfo {pages} {199} (\bibinfo {year}
		{2017})}\BibitemShut {NoStop}%
	\bibitem [{\citenamefont {Pechal}, \citenamefont {Arrangoiz-Arriola},\ and\
		\citenamefont {Safavi-Naeini}(2019)}]{pechal_safavi}%
	\BibitemOpen
	\bibfield  {author} {\bibinfo {author} {\bibfnamefont {M.}~\bibnamefont
			{Pechal}}, \bibinfo {author} {\bibfnamefont {P.}~\bibnamefont
			{Arrangoiz-Arriola}}, \ and\ \bibinfo {author} {\bibfnamefont {A.~H.}\
			\bibnamefont {Safavi-Naeini}},\ }\href@noop {} {\bibfield  {journal}
		{\bibinfo  {journal}
			{\href{https://iopscience.iop.org/article/10.1088/2058-9565/aadc6c/pdf}{Quantum\,Sci.\,Technol.}}\
		}\textbf {\bibinfo {volume} {4}},\ \bibinfo {pages} {015006} (\bibinfo {year}
		{2019})}\BibitemShut {NoStop}%
	\bibitem [{\citenamefont {Arrangoiz-Arriola}\ \emph {et~al.}(2018)\citenamefont
		{Arrangoiz-Arriola}, \citenamefont {Wollack}, \citenamefont {Pechal},
		\citenamefont {Witmer}, \citenamefont {Hill},\ and\ \citenamefont
		{Safavi-Naeini}}]{arrangoiz_PRX}%
	\BibitemOpen
	\bibfield  {author} {\bibinfo {author} {\bibfnamefont {P.}~\bibnamefont
			{Arrangoiz-Arriola}}, \bibinfo {author} {\bibfnamefont {E.~A.}\ \bibnamefont
			{Wollack}}, \bibinfo {author} {\bibfnamefont {M.}~\bibnamefont {Pechal}},
		\bibinfo {author} {\bibfnamefont {J.~D.}\ \bibnamefont {Witmer}}, \bibinfo
		{author} {\bibfnamefont {J.~T.}\ \bibnamefont {Hill}}, \ and\ \bibinfo
		{author} {\bibfnamefont {A.~H.}\ \bibnamefont {Safavi-Naeini}},\ }\href@noop
	{} {\bibfield  {journal} {\bibinfo  {journal}
			{\href{https://journals.aps.org/prx/pdf/10.1103/PhysRevX.8.031007}{Phys.\,Rev.\,X}}\
		}\textbf {\bibinfo {volume} {8}},\ \bibinfo {pages} {031007} (\bibinfo {year}
		{2018})}\BibitemShut {NoStop}%
	\bibitem [{\citenamefont {Arrangoiz-Arriola}\ \emph {et~al.}(2019)\citenamefont
		{Arrangoiz-Arriola}, \citenamefont {Wollack}, \citenamefont {Wang},
		\citenamefont {Pechal}, \citenamefont {Jiang}, \citenamefont {McKenna},
		\citenamefont {Witmer},\ and\ \citenamefont
		{Safavi-Naeini}}]{arrangoiz_arxiv}%
	\BibitemOpen
	\bibfield  {author} {\bibinfo {author} {\bibfnamefont {P.}~\bibnamefont
			{Arrangoiz-Arriola}}, \bibinfo {author} {\bibfnamefont {E.~A.}\ \bibnamefont
			{Wollack}}, \bibinfo {author} {\bibfnamefont {Z.}~\bibnamefont {Wang}},
		\bibinfo {author} {\bibfnamefont {M.}~\bibnamefont {Pechal}}, \bibinfo
		{author} {\bibfnamefont {W.}~\bibnamefont {Jiang}}, \bibinfo {author}
		{\bibfnamefont {T.~P.}\ \bibnamefont {McKenna}}, \bibinfo {author}
		{\bibfnamefont {J.~D.}\ \bibnamefont {Witmer}}, \ and\ \bibinfo {author}
		{\bibfnamefont {A.~H.}\ \bibnamefont {Safavi-Naeini}},\ }\href@noop {}
	{}\bibinfo {howpublished} {{preprint at
			\href{https://arxiv.org/pdf/1902.04681.pdf}{arXiv: 1902.04681}}} (\bibinfo
	{year} {2019})\BibitemShut {NoStop}%
	\bibitem [{\citenamefont {Sarabalis}\ \emph {et~al.}(2019)\citenamefont
		{Sarabalis}, \citenamefont {Dahmani}, \citenamefont {Cleland},\ and\
		\citenamefont {Safavi-Naeini}}]{sarabalis_arxiv}%
	\BibitemOpen
	\bibfield  {author} {\bibinfo {author} {\bibfnamefont {C.~J.}\ \bibnamefont
			{Sarabalis}}, \bibinfo {author} {\bibfnamefont {Y.~D.}\ \bibnamefont
			{Dahmani}}, \bibinfo {author} {\bibfnamefont {A.~Y.}\ \bibnamefont
			{Cleland}}, \ and\ \bibinfo {author} {\bibfnamefont {A.~H.}\ \bibnamefont
			{Safavi-Naeini}},\ }\href@noop {} {}\bibinfo {howpublished} {{preprint at
			\href{https://arxiv.org/pdf/1904.04981.pdf}{arXiv:1904.04981}}} (\bibinfo
	{year} {2019})\BibitemShut {NoStop}%
	\bibitem [{\citenamefont {Morgan}(2007)}]{morgan}%
	\BibitemOpen
	\bibfield  {author} {\bibinfo {author} {\bibfnamefont {D.}~\bibnamefont
			{Morgan}},\ }\href@noop {} {\emph {\bibinfo {title} {Surface Acoustic Wave
				Filters}}}\ (\bibinfo  {publisher} {Academic Press},\ \bibinfo {year}
	{2007})\BibitemShut {NoStop}%
	\bibitem [{\citenamefont {Hashimoto}(2009)}]{hashimoto}%
	\BibitemOpen
	\bibfield  {author} {\bibinfo {author} {\bibfnamefont {K.-Y.}\ \bibnamefont
			{Hashimoto}},\ }\href@noop {} {\emph {\bibinfo {title} {RF Bulk Acoustic Wave
				Filters for Communications}}}\ (\bibinfo  {publisher} {Artech House},\
	\bibinfo {year} {2009})\BibitemShut {NoStop}%
	\bibitem [{\citenamefont {Pop}\ \emph {et~al.}(2017)\citenamefont {Pop},
		\citenamefont {Kochhar}, \citenamefont {Vidal-Alvarez},\ and\ \citenamefont
		{Piazza}}]{pop_piazza}%
	\BibitemOpen
	\bibfield  {author} {\bibinfo {author} {\bibfnamefont {F.~V.}\ \bibnamefont
			{Pop}}, \bibinfo {author} {\bibfnamefont {A.~S.}\ \bibnamefont {Kochhar}},
		\bibinfo {author} {\bibfnamefont {G.}~\bibnamefont {Vidal-Alvarez}}, \ and\
		\bibinfo {author} {\bibfnamefont {G.}~\bibnamefont {Piazza}},\ }\bibfield
	{title} {\enquote {\bibinfo {title} {Laterally vibrating lithium niobate
				{MEMS} resonators with 30\% electromechanical coupling coefficient},}\ }in\
	\href@noop {} {\emph {\bibinfo {booktitle}
			{\href{https://ieeexplore.ieee.org/stamp/stamp.jsp?tp=&arnumber=7863571}{Proc.
					IEEE 30th Int. Conf. Micro-Elect. Mech. Syst. (MEMS)}}}}\ (\bibinfo {year}
	{2017})\ p.\ \bibinfo {pages} {966}\BibitemShut {NoStop}%
	\bibitem [{\citenamefont {Lakin}(1992)}]{lakin}%
	\BibitemOpen
	\bibfield  {author} {\bibinfo {author} {\bibfnamefont {K.~M.}\ \bibnamefont
			{Lakin}},\ }\bibfield  {title} {\enquote {\bibinfo {title} {Modeling of thin
				film resonators and filters},}\ }in\ \href@noop {} {\emph {\bibinfo
			{booktitle}
			{\href{https://ieeexplore.ieee.org/stamp/stamp.jsp?tp=&arnumber=187931}{Proc.
					IEEE MTT-S Int. Microwave Symp. Dig.}}}}\ (\bibinfo {year} {1992})\ p.\
	\bibinfo {pages} {149}\BibitemShut {NoStop}%
	\bibitem [{\citenamefont {Pozar}(2012)}]{pozar}%
	\BibitemOpen
	\bibfield  {author} {\bibinfo {author} {\bibfnamefont {D.~M.}\ \bibnamefont
			{Pozar}},\ }\href@noop {} {\emph {\bibinfo {title} {Microwave Engineering}}}\
	(\bibinfo  {publisher} {Wiley},\ \bibinfo {year} {2012})\BibitemShut
	{NoStop}%
	\bibitem [{\citenamefont {Schuster}(2007)}]{schuster_thesis}%
	\BibitemOpen
	\bibfield  {author} {\bibinfo {author} {\bibfnamefont {D.~I.}\ \bibnamefont
			{Schuster}},\ }\emph {\bibinfo {title} {Circuit Quantum Electrodynamics}},\
	\href@noop {} {\bibinfo {type} {{Ph.D.} thesis}},\ \bibinfo  {school} {Yale
		University} (\bibinfo {year} {2007})\BibitemShut {NoStop}%
	\bibitem [{\citenamefont {Reed}(2013)}]{reed_thesis}%
	\BibitemOpen
	\bibfield  {author} {\bibinfo {author} {\bibfnamefont {M.~D.}\ \bibnamefont
			{Reed}},\ }\emph {\bibinfo {title} {Entanglement and Quantum Error Correction
			with Superconducting Qubits}},\ \href@noop {} {\bibinfo {type} {{Ph.D.}
			thesis}},\ \bibinfo  {school} {Yale University} (\bibinfo {year}
	{2013})\BibitemShut {NoStop}%
\end{thebibliography}
\end{document}